# LLMs in Qualitative Research: Opportunities, Limitations, and Practical Considerations

# LLMs in Qualitative Research: Opportunities, Limitations, and Practical Considerations

## Introduction

This full methods paper examines the opportunities, limitations, and practical considerations associated with the use of large language models (LLMs) in qualitative research. LLMs such as *Claude, Gemini, Grok, DeepSeek, and ChatGPT* have rapidly become prominent systems in both public and academic discourse. In fact, OpenAI, the parent company of ChatGPT, reported approximately 700 million weekly active users as of September 2025 [1]. Alongside this rapid growth, however, important questions have emerged regarding how these models are developed and deployed. These concerns include the scraping of copyrighted materials for training data, the environmental costs associated with large-scale data centers, particularly water and energy consumption by Graphics Processing Units (GPUs), and broader issues related to data privacy [2], [3], [4], [5]. Many widely used LLMs are also proprietary, which limits accessibility around their training data, model weights, and architectures [6].

It is within the context of this broader ethical and social debate that our interdisciplinary research team seeks to explore the use of LLMs in qualitative research. Importantly, this paper is authored by a multidisciplinary team comprising engineering education researchers with extensive experience in qualitative methods and computer scientists with expertise in explainable AI. We note that while contemporary LLMs differ in their parameters, training data, and licensing models, they share common design considerations that we focus on here, that is, a curated set of features/hyperparameters directly relevant to the quality and trustworthiness of qualitative analysis and most accessible to researchers. This set includes context window limitations, temperature settings (a parameter that controls how deterministic or varied the model's responses are), user/system prompts, and other considerations that cut across contemporary models and are likely to remain relevant regardless of which specific model is used. By discussing both the opportunities and limitations associated with these features, we aim to offer awareness for qualitative researchers who may be considering the use of LLMs in their work.

## Background

### *Natural Language Processing and Qualitative Research*

Natural language processing methods have long served as a supporting tool for qualitative researchers. Earlier approaches, such as topic modeling (e.g., Latent Dirichlet Allocation) [7] and lexicon-based sentiment analysis models [8], were designed to assist researchers in organizing and exploring large volumes of textual data [9]. However, these methods generally provided researchers with substantial procedural transparency. For example, preprocessing decisions (e.g., stop word removal), representational assumptions (e.g., bag-of-words), and hyperparameters (e.g., number of topics) were explicitly specified and directly adjustable. Although these models were statistical abstractions, their underlying mechanics were comparatively *explainable*, meaning researchers could trace how word frequencies and co-occurrence patterns contributed to outputs. For instance, a researcher using LDA to analyze transcripts of faculty interviews could inspect the terms driving each topic, remove words that were frequent but conceptually uninformative (e.g., "the," "said," "think"), and adjust the number of topics, all guided by their own domain knowledge of the subject matter.

Contemporary LLMs, by contrast, operate on a fundamentally different scale. While we note that open-source LLMs offer greater access to source code, architectures, and training procedures, they still involve billions or trillions of parameters, rendering their internal decision-making processes generally opaque [10]. This opaque nature raises concerns about trustworthiness, particularly given longstanding evidence that even open-source models can learn and reproduce biases present in training data [11], [12], [13], [14]. These concerns are compounded by recent findings showing that even state-of-art *reasoning* models may show users steps that do not align with their internal computations [15]. Although active research in mechanistic and explainable AI continues to advance understanding of neural network's internal mechanisms (e.g., neurons activations, encoded representations, attention heads, causal pathways) [16], [17], [18], significant gaps remain between current interpretability methods and the level of procedural access available with earlier approaches [10]. For a novice qualitative researcher, these gaps mean that adopting LLMs for qualitative analysis requires engaging with a set of methodological considerations that are different from those associated with earlier NLP tools, and that the remainder of this paper aims to make accessible.

### *Why Qualitative Research Requires Special Consideration*

Beyond these technical concerns, we argue that the suitability of LLMs depends not only on the possible efficiency gains but also on how their use aligns with disciplinary norms, epistemological assumptions, and standards of quality. These considerations are especially relevant in qualitative research, where meaning-making and interpretation are central to knowledge production [19].

Qualitative research is characterized by the systematic analysis of rich, contextualized textual data, often collected through interviews, ethnographies, focus groups, and open-ended questions. For example, qualitative researchers may examine *why* individuals interpret experiences in certain ways, or how social, institutional, and cultural contexts influence the construction of meaning. Answering these questions requires careful attention to nuance, context, and participant voice. Ensuring the trustworthiness of qualitative findings typically involves iterative and labor-intensive processes, including prolonged engagement with data, reflexive memoing, collaborative coding of data, and triangulation across data sources and researchers [20]. Importantly, interpretation is not treated as a purely technical step but as an analytic act shaped by theory and researchers' lived experiences. As a result, qualitative researchers are expected to make their perspectives explicit and to reflect on how these perspectives influence analytic decisions. This reflexive practice is often articulated through positionality statements [21].

Additionally, unlike many positivist approaches, qualitative research does not generally aim for statistical generalizability [22]. Instead, it seeks to produce rich descriptions of specific contexts that enable readers to assess the potential *transferability* of findings to other settings. These epistemological commitments make qualitative research particularly sensitive to automation. In fact, many qualitative researchers have raised concerns about the potential for LLMs to displace the distinctly human dimensions of meaning-making, including reflection, positionality, and interpretive judgment [23], we share these concerns, but also remain open-minded about potential stages of research that LLMs could help in.

Our work is grounded in constructivism, which assumes that meaning is co-constructed through engagement between researchers and data [22]. We therefore do not think of LLM outputs as objective interpretations, but as products that enter the researcher's interpretive process, where reflexive evaluation and judgment remain with the researcher. We also acknowledge that LLMs do not hold positionality in the way a researcher does [24], but also note that models trained predominantly on data may reflect certain cultural norms and these may implicitly privilege certain experiences while marginalizing others [25].

As such, scholars have begun to address these concerns by proposing quality frameworks to guide the responsible use of LLMs in research. For example, Reeping et al. [24] applied the *Qualifying Qualitative Research Quality Framework* by [26] to evaluate LLM integration into reflective engineering education research. Other work has also explored frameworks such as *Inference to the Best Explanation* to assess the validity of model explanations [27]. We note that further work is needed to develop and validate quality frameworks that can guide the trustworthy integration of AI models across the various stages of the research cycle.

### *LLMs in Qualitative Research*

Scholars have already begun experimenting with LLMs across multiple stages of qualitative research, from initial text exploration to thematic analysis. Table 1 provides an overview of a general research process and highlights ways in which LLMs have been used to support different stages of this process. In the area of text clustering, Grootendorst et al. [28] developed a topic modeling approach that leverages transformer-based embeddings to create dense clusters of semantically similar text, producing easily interpretable topics. Gao et al. [29] proposed a workflow in which AI-generated coding suggestions supported consensus building among researchers, while Xiao et al. [30] explored deductive coding through systematic prompt design and reported agreement with expert-coded results. Anakok et al. [31] examined the use of LLMs to the multiple stages of thematic analysis, including data familiarization, coding, theme generation, refinement, and write-up. Similarly, De Paoli [32] found that LLMs can perform inductive thematic analysis with a reasonable degree of validity and Katz et al. [33] developed the Generative AI enabled Theme Organization and Structuring (GATOS) workflow to support aspects of thematic analysis, notably using open-source models rather than proprietary models. Additionally, Ross and Katz [34] further demonstrated the usage of GATOS to conduct thematic analysis on a large amount of Reddit posts while maintaining human-in the loop process. In terms of efficiency, Quia [35] found that using LLMs for inductive coding reduced their coding time by 81% while maintaining 80% accuracy when applied to coding semi-structured interviews.

However, comparative studies reveal that LLM performance is inconsistent across contexts. Several scholars have compared LLM-assisted analysis with manual coding approaches and reported results suggesting that LLMs can produce outputs that are, in some cases, comparable to humans [36],[37],[38]. Yet performance appears sensitive to both model version and task characteristics, for example Burgess et al. study found that more recent versions of models showed better agreement with human coders [39], while another found that coding accuracy decreased as document length increased [40]. This variability in LLM performance across qualitative tasks is also a motivation for the present paper; By describing several factors that shape model outputs, we aim to help researchers make more informed methodological decisions when integrating LLMs into their work.

**Table 1: LLMs in research**

| Example Task | How LLMs could be used | Representative Work |
|---|---|---|
| Literature search | Finding relevant papers, summarizing articles | [41] |
| Research design & protocol development | Drafting and refining protocols (e.g., interview guides, survey items) | [42], [43] |
| Interview practice | Simulating participant responses | [44] |
| Transcription | Translating multilingual interviews, cleaning interviews | [45] |
| Text exploration & organization | Creating interpretable topic clusters from qualitative data | [28] |
| Coding and codebook development | Generating candidate codes | [46] |
| Thematic analysis | Supporting familiarization, coding, theme generation, and refinement | [31] |
| Human vs. LLM comparison / validation | Comparing LLM outputs to human coding for alignment, reliability, or efficiency | [47] |

## Technical Considerations for Qualitative Researchers

The following section covers technical considerations, specifically, a curated set of parameters and design choices that qualitative researchers should be aware of when working with LLMs. These are not exhaustive but reflect the factors most likely to influence the quality and reproducibility of LLM-assisted analyses. To support practical adoption, we also provide a companion GitHub repository (https://github.com/eigenhenry/asee_large_models) containing a sample workflow that illustrates how these considerations can be implemented programmatically.

### *Review System Cards*

System cards (also known as model cards) describe a model's intended use, training data characteristics, safeguards, known limitations, and evaluation practices [48], [49]. For qualitative researchers, system cards could be an important resource for reflexive and responsible use. They provide insight into whose perspectives may be overrepresented in training data, which forms of content are filtered, and where known model weaknesses lie. Reading and documenting system-level instructions and model documentation should therefore be treated as an integral part of the qualitative research process, akin to reflecting positionality, analytic assumptions, and methodological constraints. Researchers can find system cards on the respective website of the LLM of interest. Unfortunately, system cards vary considerably in their depth and completeness across providers, and some models lack them entirely.

### *Open-source or Proprietary Models?*

While proprietary models (e.g., Gemini, Claude) remain widely used, particularly through consumer-facing interfaces, open-source models are gaining momentum [50]. Major industry players have contributed to this trend by releasing openly available models, including the GPT OSS model from OpenAI and the LLaMA models from Meta. Open-source models are generally free to use and offer researchers greater access to training data, architectures, and model weights. This transparency can provide meaningful advantages for research, including greater control over data privacy and the ability to inspect and modify model behavior [6]. However, many models marketed as *open* often only release their weights while keeping training data, curation procedures, and other details of system development proprietary [6]. Additionally, while open-source models offer greater transparency in principle, deploying them locally typically requires significant computational resources which may not be available to all researchers. For example, deploying even a relatively small model such as Qwen-8B requires approximately 8 GB of RAM merely to load its weights, with additional memory necessary for inference.

### *Programmatic Options for Standardized Workflows*

Regardless of whether researchers use proprietary or open-source models, the mode of interaction also shapes the degree of methodological control available to them. Chat-based interfaces, while accessible, often obscure important hyper parameters (e.g., temperature) that can systematically structure their analyses. Programmatic interaction enables researchers to implement standardized, repeatable workflows in which inputs are systematically batched, prompting is held constant across cases, and outputs are archived in structured formats that support transparency, comparison, and analytic reexamination. For example, a researcher could apply the same prompt to each transcript, hold the temperature and system prompt constant across all cases, and export the model's responses into a structured format such as a spreadsheet for subsequent analysis, see GitHub (https://github.com/eigenhenry/asee_large_models) for an example on how to use programmatic access.

### *Be Careful with the Context Window*

Another key concept underlying LLM-based workflows is the context window, sometimes described as the model's *working memory*. The context window defines how much text a model can consider at one time when generating a response, measured in units called tokens. For practical purposes, a token corresponds roughly to four characters of English text, meaning that a single word is often one token, though common words may be split differently depending on the model. Contemporary models vary widely in this capacity, with some supporting from 32k tokens to others claiming capacities in the millions. Importantly, when the number of tokens provided exceeds the context window, earlier portions of the input are truncated or ignored. Moreover, the advertised context window does not represent the full usable capacity as different tokenizers segment text differently across models, and a portion of the window is typically consumed by system-level instructions and conversation history [51].

These constraints have direct implications for qualitative research. Many researchers interact with LLMs through chat-based interfaces and may be unaware when the context window has been exceeded. A researcher attempting to upload dozens of interview transcripts, for example, may find that the model drops portions of the data, leading to incomplete or misleading analyses. Prior work in computer science has also shown that model performance can degrade as input length increases, even when the text technically fits within the window [40]. This variation has also been documented, for example, in multilingual research, where the same text translated into different languages can produce drastically different token counts [52]. Researchers can estimate the length of their inputs using tools such as *Tokenizer Playground* [53], which provide approximate token counts for a given text and model.

### *Adjust Temperature and/or Top-p values*

Non-determinism arises from the fact that LLMs do not produce a single fixed output for a given prompt. Instead, they sample from a probability distribution over possible next-token sequence. Two key mechanisms for controlling this behavior are the temperature and top-p (nucleus sampling) hyper-parameters. Temperature, which typically ranges from 0 to 1, adjusts the shape of the probability distribution itself, lower temperature values bias the model toward selecting the most probable next token, resulting in more consistent and predictable outputs [54]. Higher temperature values introduce greater randomness, producing more varied and creative responses. Top-p operates differently by limiting the pool of tokens the model can sample from; A top-p value of 0.9, for example, restricts selection to the smallest set of tokens whose cumulative probability reaches 90%, excluding less likely options. Lower top-p values narrow this pool further, producing more focused outputs, while higher values allow the model to draw from a broader range of possibilities. In practice, temperature and top-p interact, and therefore should be carefully adjusted independently [55], [56]. For qualitative research, temperature settings have direct methodological implications. Tasks that require consistency, such as applying codes or identifying themes across a dataset, may benefit from a temperature of zero or near zero as done in [33].

In contrast, exploratory tasks such as brainstorming analytic questions or generating alternative interpretations may benefit from higher temperature values. However, increased randomness could also raise the likelihood of hallucinations, particularly for tasks that require factual accuracy or close alignment with the data [57]. Importantly, many chat-based interfaces default to intermediate temperature values (often around 0.5) and do not clearly expose this setting. Additionally, even when temperature is set to zero, many commercial LLM APIs do not guarantee fully deterministic outputs due to floating-point non-determinism inherent in GPU computation. In other words, identical inputs submitted under identical parameter settings may still produce slightly different outputs across GPU runs. For qualitative researchers concerned with procedural rigor and the auditability of analytic decisions, this residual variability should be acknowledged and, where possible, mitigated by archiving all model outputs alongside the prompts and parameter settings used to generate them.

***Prompt Techniques and System Prompts can Shape Outputs***

While researchers are increasingly aware of the role of prompt techniques (e.g., chain-of-thought prompting) in shaping LLM outputs and performance [58], the role of system prompts deserves equal attention. System prompts are instructions typically provided by model developers that guide model behavior across all user interactions, shaping what types of questions a model will answer and how it frames responses. For example, proprietary models often include restrictions on generating content related to certain sensitive topics, regardless of the researcher's intent. These constraints can shape analytic outputs in ways that are not always visible to users. System prompts can be intentionally leveraged by researchers to support specific qualitative tasks. A researcher may, for instance, define analytic roles or personas through system prompts to simulate interview protocols or examine how prompting variations shape narrative responses. Importantly, both user and system prompts should employ clear, unambiguous language and explicit instructions, for more information on how to structure prompts see [59]. Table 2 summarizes the key technical considerations discussed in this section and their practical implications for qualitative researchers.

**Table 2:** Practical Considerations for Using LLMs in Qualitative Research

| Consideration | What It Is | Risk for Qualitative Research | Example Considerations |
|---|---|---|---|
| **Model Selection** | Choice between proprietary (e.g., Claude) and open source/weight (e.g., Qwen) | Proprietary models limit transparency; running open-source models may require hardware not available to all researchers | Consider data privacy needs / available resources |
| **Mode of interaction** | Chat-based vs programmatic interface | Chat interfaces obscure hyperparameters / programmatic access requires coding knowledge | Consider when there's need of batching documents (due to token count) or need to adjust parameters (e.g. temperature) |
| **Context Window** | Maximum text a model can process at once | Model may not consider all input context if exceeded, performance for some tasks degrade with lengths | Estimate token counts using model-specific tokenizers; segment large corpora of text |
| **Temperature & Top p** | Hyperparameters controlling output randomness | Higher values increase variability; lower values are more deterministic | Set low values for consistent dependent tasks, high values for more "creative" tasks |
| **User and System prompts** | Developer or user-defined instructions shaping model behavior and output | Proprietary restrictions may invisibly constrain outputs / system prompts can be set by researcher in programmatic access | Document all system prompts used; leverage custom prompts to define analytic roles |
| **System cards** | Documentation describing data, limitations, and safeguards | Without review, researchers may be unaware of possible worldviews via content filtering | Read and document system cards as part of the research design process |

## Limitations

This paper has several limitations. First, we focused on a curated set of technical parameters most accessible to qualitative researchers but did not address other considerations such as retrieval-augmented generation (RAG), fine-tuning, or multi-modal capabilities that may also shape research outputs. Second, the LLM landscape evolves rapidly, and specific details discussed here, such as context window sizes and default parameter values, may shift as models are updated. While we focused on concepts likely to remain relevant across systems, practical details should be verified at the time of use. Third, this paper synthesizes existing literature and offers practical guidance but does not include original empirical data demonstrating how adjusting these parameters affects qualitative research outcomes. Finally, while we reference multilingual tokenization differences, our discussion is primarily situated within English-language qualitative research in engineering education, and researchers working in other languages or disciplines may face additional challenges that this paper does not fully address.

## Implications and Conclusion

While we, the authors, agree that researchers should continue to lead the reflective and interpretive dimensions of qualitative work, the literature suggests that LLMs can meaningfully support other stages of the research process. Tasks such as data familiarization, transcript cleaning, and interview protocol practice may benefit from LLM assistance, and for tasks that do not require the same degree of human sensitivity, LLMs could free researchers to focus more of their time on the analytic work that matters most. There is also potential for developing structured workflows that integrate LLMs into qualitative research, whether building on existing approaches such as GATOS [33] or using LLM outputs as an additional point of triangulation alongside human analysis [42], [42].

LLM performance is likely to vary depending on the model, parameters, and mode of interaction used. We argue that by understanding the concepts described in this paper, including temperature, top-p, context windows, system prompts, and system cards, researchers gain greater agency over the outputs produced by models. Similarly, understanding the differences between proprietary and open-source models allows researchers to make more informed decisions based on their computational resources, privacy needs, and transparency requirements. Additionally, while it is common in computer science to document these decisions, such practices are not yet standard in engineering education research [60]. We therefore suggest that these parameters be documented as part of a study's methodological reporting, much as researchers currently report their coding procedures, inter-rater processes, and analytic frameworks.

More broadly, we believe that this awareness positions researchers to engage with LLMs collaboratively rather than passively. When researchers understand how outputs are shaped, they are better prepared to make principled decisions about where LLMs can meaningfully support qualitative work and where human judgment must remain central. This critical engagement could, in turn, support the development of more robust frameworks for integrating LLMs into qualitative research, frameworks grounded in disciplinary norms of quality, reflexivity, and researcher autonomy rather than in the default configurations of any particular model.

## Use of Generative AI

Generative AI tools were used to assist with literature review of this paper. Specifically, we used Scholar Labs and ChatGPT's Deep Research to help find relevant papers. All found papers were read thoroughly by the authors to ensure accuracy and integrity.